\documentclass[onecolumn,12pt,tightenlines,superscriptaddress,notitlepage]{revtex4-1}

\usepackage{amsmath}
\usepackage{amssymb}
\usepackage{latexsym}

\usepackage{url}
\usepackage{xcolor}

\usepackage{graphicx}
\definecolor{newcolor}{rgb}{.8,.349,.1}

\date{\today}

\newcommand{\dd}[2]{\frac{\mathrm{d}{#1}}{\mathrm{d}{#2}}}
\newcommand{\pd}[2]{\frac{\partial{#1}}{\partial{#2}}}

\begin{document}

\begin{abstract}

Numerical modeling of electromagnetic waves is an important tool for understanding the interaction of light and matter, and lies at the core of computational electromagnetics. Traditional approaches to injecting and evolving electromagnetic waves, however, can be prohibitively expensive and complex for emerging problems of interest and can restrict the comparisons that can be made between simulation and theory. As an alternative, we demonstrate that electromagnetic waves can be incorporated analytically by decomposing the physics equations into analytic and computational parts. In particle-in-cell simulation of laser--plasma interaction, for example, treating the laser pulse analytically enables direct examination of the validity of approximate solutions to Maxwell's equations including Laguerre--Gaussian beams, allows lower-dimensional simulations to capture 3-D--like focusing, and facilitates the modeling of novel space--time structured laser pulses such as the flying focus. The flexibility and new routes to computational savings introduced by this analytic pulse technique are expected to enable new simulation directions and significantly reduce computational cost in existing areas.

\end{abstract}


\title{Analytic pulse technique for computational electromagnetics}

\author{K. Weichman}
\email[corresponding author, ]{kweic@lle.rochester.edu}
\affiliation{Laboratory for Laser Energetics, University of Rochester, Rochester, NY 14623, USA}

\author{K. G. Miller}
\affiliation{Laboratory for Laser Energetics, University of Rochester, Rochester, NY 14623, USA}

\author{B. Malaca}
\affiliation{GoLP/Insituto de Plasmas e Fus\~ao Nuclear, Instituto Superior T\'ecnico, Universidade de Lisboa, Lisbon 1049-001, Portugal}

\author{W. B. Mori}
\affiliation{Department of Physics and Astronomy, University of California, Los Angeles, CA 90095, USA}

\author{J. R. Pierce}
\affiliation{Department of Physics and Astronomy, University of California, Los Angeles, CA 90095, USA}

\author{D. Ramsey}
\affiliation{Laboratory for Laser Energetics, University of Rochester, Rochester, NY 14623, USA}

\author{J. Vieira}
\affiliation{GoLP/Insituto de Plasmas e Fus\~ao Nuclear, Instituto Superior T\'ecnico, Universidade de Lisboa, Lisbon 1049-001, Portugal}

\author{M. Vranic}
\affiliation{GoLP/Insituto de Plasmas e Fus\~ao Nuclear, Instituto Superior T\'ecnico, Universidade de Lisboa, Lisbon 1049-001, Portugal}

\author{J. P. Palastro}
\affiliation{Laboratory for Laser Energetics, University of Rochester, Rochester, NY 14623, USA}

\maketitle

\section{Introduction} 

The numerically accurate representation of electromagnetic waves is a critical component of modeling many physical systems. 
Despite over 60 years of research in computational electromagnetics, 
the modeling of electromagnetic waves remains an active subject, motivated by the ever-increasing diversity and demands of light--matter interaction physics. 
For example, emerging areas of laser--plasma interaction and nonlinear optics, such as tightly focused beams~\cite{thiele2016tight,jirka2017pair},
strong-field quantum electrodynamics (QED)~\cite{blackburn2017pair,amaro2021pair},
and laser pulses with spatiotemporal structure~\cite{kondakci2017spatio,sainte-marie2017spatio,froula2018ff,palastro2020dephasingless,pierce2023astrl} present new computational challenges.
In each of these areas, the development of new simulation techniques is driven by the high cost, high complexity, or both, of the traditional approaches to injecting and advancing laser pulses in grid-based electromagnetic simulations.

In grid-based methods, the electromagnetic fields are evolved according to a cell-based discretization of Maxwell's equations, with boundary conditions and laser--matter coupling terms dictated by the problem being considered.
Electromagnetic waves, e.g., laser pulses, are traditionally incorporated either via initialization of the wave packet in its entirety in the simulation domain or through numerical antennae.
In many cases, traditional techniques are successful and appropriate, such as when the laser is not overly burdensome to introduce into the domain, when the dimensionality of current evolution matches the dimensionality of laser focusing, and when approximations to the laser pulse profile used to formulate theory are valid. 

Emerging areas of interest, however, can depart from these conditions and require advanced techniques. 
For instance, the tightly focused beams commonly employed in high-field laser--plasma interactions~\cite{sokolov2010compton,bulanov2013compton} 
are inconsistent with the paraxial approximation frequently used in theoretical models. The effects of departure from the paraxial approximation may be important~\cite{popov2008tight} but cannot be investigated directly in simulations that only utilize full solutions to Maxwell's equations. 
In addition, the requirement for simulations to be fully 3-D in order to reproduce 3-D laser focusing introduces a significant computational expense that is unnecessary when the dynamics of particles are lower-dimensional, for example in certain QED processes, such as inverse Compton scattering from high-energy electron beams~\cite{blackburn2017pair,amaro2021pair}. 
Traditional methods, especially those resolving the laser wavelength, can also be prohibitive for novel laser pulses. For spatiotemporally shaped flying-focus pulses~\cite{kondakci2017spatio,sainte-marie2017spatio,froula2018ff,palastro2020dephasingless}, capturing interesting dynamics may require simulating many Rayleigh ranges, leading to either very large (and expensive) simulations or complicated (and yet-to-be developed) boundary conditions.
Lastly, the modeling of pulsed electromagnetic waves often requires incoming-wave-only boundary conditions, which are simplest to formulate in vacuum and are best-behaved for a single pre-specified angle of incidence.
Departure from these conditions traditionally necessitates a more-complex treatment, such as padding the physical domain with absorbing numerical regions~\cite{berenger1994pml,merewether1980tfsf,schneider2006tfsf,tan2010tfsf,capoglu2013tfsf}.

In this paper, we introduce a new, half-analytic approach to incorporating electromagnetic pulses in simulations using analytic solutions to Maxwell's equations. 
This analytic pulse technique (APT) produces identical results to unmodified simulation algorithms when numerical effects, e.g., dispersion, are properly matched (Sec.~\ref{sec:theory}). 
Using the example of finite-difference time domain particle-in-cell simulation, we demonstrate that APT exhibits good performance (Sec.~\ref{sec:performance}) and addresses cost- and complexity-related limitations of the traditional approach for emerging areas of laser--plasma interaction (Sec.~\ref{sec:applications}). For example, the modified field update enables direct evaluation of approximate solutions to Maxwell's equations (Sec.~\ref{sec:paraxial}); allows 3-D laser intensity profiles to be included in lower-dimensionality simulations (Sec.~\ref{sec:dims}); and facilitates consideration of novel, otherwise prohibitively expensive or complex laser pulses (Sec.~\ref{sec:ff}). 
In addition, an analytic current can significantly simplify the treatment of plasma-filled domains by cancelling the physical current at the simulation boundary (Sec.~\ref{sec:j1}). 
These advantages are expected to enable new simulation directions and significantly reduce computational cost in existing areas, making the modified field update an attractive approach to modeling laser pulses in computational electromagnetics.

\section{Formulation of the analytic pulse technique} \label{sec:theory}

In computational modeling of light--matter interactions, electromagnetic waves are coupled to the material response via Maxwell's equations or one of its variants (e.g., the unidirectional pulse-propagation equation~\cite{uppe2002kolesik}). In electromagnetic plasma simulations, for example, solvers for the electric and magnetic fields are coupled to solvers for the current.
The fields are governed by a numerical discretization of Maxwell's equations,
\begin{align}
    & \dfrac{1}{c}\pd{\mathbf{B}}{t} = - \nabla \times \mathbf{E} \\
    & \dfrac{1}{c}\pd{\mathbf{E}}{t} = \nabla \times \mathbf{B} - \dfrac{4\pi}{c} \mathbf{J}. \label{eqn:b_maxwell}
\end{align}
Without loss of generality, it is possible to decompose the electric and magnetic fields in the above equations into two parts $\mathbf{E} = \mathbf{E_1}+ \mathbf{E_2}$ and $\mathbf{B} = \mathbf{B_1}+ \mathbf{B_2}$, such that $\mathbf{E_1}$ and $\mathbf{B_1}$ satisfy Maxwell's equations in vacuum, and the contribution to the fields from the plasma is entirely encoded in $\mathbf{E_2}$ and $\mathbf{B_2}$,
\begin{align}
    & \dfrac{1}{c}\pd{\mathbf{B_1}}{t} = - \nabla \times \mathbf{E_1} \;\; ; \;\; \dfrac{1}{c}\pd{\mathbf{E_1}}{t} = \nabla \times \mathbf{B_1} \label{eqn:E_1}\\
    & \dfrac{1}{c}\pd{\mathbf{B_2}}{t} = - \nabla \times \mathbf{E_2} \;\; ; \;\; \dfrac{1}{c}\pd{\mathbf{E_2}}{t} = \nabla \times \mathbf{B_2} - \dfrac{4\pi}{c} \mathbf{J} \label{eqn:E_2}, 
\end{align}
where $\mathbf{J}$ depends on the total fields $\mathbf{E_1}+ \mathbf{E_2}$ and $\mathbf{B_1}+ \mathbf{B_2}$. The simulation can thereby be split to advance the electric and magnetic fields as they would be in vacuum (subscript $1$, the ``vacuum fields'') separately from the fields generated by the plasma response (subscript $2$).

The approaches to updating the vacuum fields and the plasma response do not, in general, need to be the same. Algorithms that produce the same discrete solution to Maxwell's equations will produce the same total electric and magnetic fields and result in identical plasma dynamics. Therefore, $\mathbf{E_1}$ and $\mathbf{B_1}$ can be set using an approximate analytic solution to Maxwell's equations, while $\mathbf{E_2}$ and $\mathbf{B_2}$ are evolved with the usual field solver.

A similar decomposition of the fields into analytic and computational parts may be applied in any simulation method in which linear terms appear in the field equations.  
While the present work will detail the implementation of the analytic pulse technique in particle-in-cell simulations, extension to methods including magnetization and polarization is straightforward [e.g., $\mathbf{J}\to \mathbf{J} + c\nabla \times \mathbf{M} + (\partial \mathbf{P}/\partial t)$, where $\mathbf{J}$, $\mathbf{M}$, and $\mathbf{P}$ are evaluated using the total fields]. 
In addition, although Eqs.~\ref{eqn:E_1} and~\ref{eqn:E_2} represent the separation between the vacuum fields and the plasma response, APT can also include analytic current terms. Analytic current can be used, for example, to cancel the physical current on the simulation boundary, which can be desirable from an algorithmic standpoint. For simplicity, the use of analytic current is restricted to Sec.~\ref{sec:j1}, whereas the remainder of the work will consider the vacuum-like case.  
In both cases, it will be shown that the analytic pulse technique applied to an electromagnetic wave, with appropriately chosen functional form, is in excellent agreement with an all-internal electromagnetic field update and can provide significant computational savings.

\subsection{Implementation in particle-in-cell simulation} \label{sec:example}

\begin{figure}
    \centering
    \includegraphics[width=0.6\linewidth]{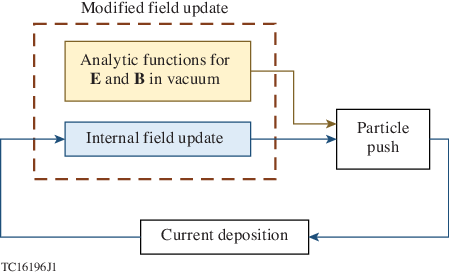}
    \caption{
    Modified PIC loop for the analytic pulse technique. In the proposed scheme, analytic functions provide an approximate solution to the discretized form of Maxwell's equations in vacuum used in the usual internal field update. The analytic fields are then added to the internally advanced fields to give the total $\mathbf{E}$ and $\mathbf{B}$ employed in the particle push. The resulting current is then used for the internal field update.}
    \label{fig:loop}
\end{figure}

To demonstrate the utility of the analytic pulse technique, the split field update described by Eqs.~\ref{eqn:E_1} and~\ref{eqn:E_2} was implemented in the finite-difference time domain particle-in-cell (PIC) algorithm. The modified PIC algorithm, shown schematically in Fig.~\ref{fig:loop}, is as follows. First, an approximate analytic solution to the discretized set of Maxwell's equations in vacuum is obtained. The resulting expressions are then used to set $\mathbf{E_1}$ and $\mathbf{B_1}$ on the grid at every time step. Next, the analytic fields and the internally-advanced fields ($\mathbf{E_2}$ and $\mathbf{B_2}$) are added together. The total fields are then interpolated to the particle position and used for the particle push. Finally, the particle current is interpolated onto the grid and is used to update the internally-advanced fields to the next time step. The success of this technique relies on finding a sufficiently accurate analytic solution, the details of which depend on the choice of field solver.

For example, consider the vacuum evolution of a monochromatic plane wave under the second-order explicit Yee scheme. 
The wave is described by
\begin{align}
& \mathbf{E} = \dfrac{1}{2}\mathbf{E_0} \, e^{i\mathbf{k}\cdot \mathbf{x} - i \omega t} + \mathrm{c.c.} \\
& \mathbf{B} = \dfrac{1}{2}\mathbf{B_0} \, e^{i\mathbf{k}\cdot \mathbf{x} - i \omega t} + \mathrm{c.c.},
\end{align}
where $\mathbf{\hat{k}}$ is perpendicular to both $\mathbf{E}$ and $\mathbf{B}$, with $\mathbf{E} \perp \mathbf{B}$. The numerical discretization of Maxwell's equations then determines the relationship between $E_0$, $B_0$, $\omega$, and $\mathbf{k}$. In what follows, $\mathbf{\hat{k}} = \mathbf{\hat{x}}$, $\mathbf{E} || \mathbf{\hat{y}}$, and $\mathbf{B} || \mathbf{\hat{z}}$. 

In the second-order explicit Yee scheme~\cite{yee1966yee}, the fields of the plane wave advance from one particle-push step to the next according to
\begin{align}
    & B_{i+1/2}^{n+1/2} = B_{i+1/2}^{n} - \dfrac{c\Delta t}{2 \Delta x}\left(E_{i+1}^{n}-E_{i}^{n}\right) \\
    & E_{i}^{n+1} = E_{i}^{n} - \dfrac{c\Delta t}{\Delta x}\left(B_{i+1/2}^{n+1/2}-B_{i-1/2}^{n+1/2}\right) \\
    & B_{i+1/2}^{n+1} = B_{i+1/2}^{n+1/2} - \dfrac{c\Delta t}{2 \Delta x}\left(E_{i+1}^{n+1}-E_{i}^{n+1}\right),
\end{align}
where $n$ is the time step index (step size $\Delta t$) and $i$ is the cell-centered index in the $x$ direction (step size $\Delta x$). This leads to the well-known 1-D dispersion relation
\begin{align} \label{eqn:dispersion}
    \sin^2\left(\dfrac{\omega \Delta t}{2}\right) = \dfrac{c^2\Delta t^2}{\Delta x^2}\sin^2\left(\dfrac{k\Delta x}{2}\right),
\end{align}
as well as
\begin{align} \label{eqn:e_amp_full}
    B_0 = \cos\left(\dfrac{\omega \Delta t}{2}\right)E_0.
\end{align}
To second order in $k \Delta x$ and $\omega \Delta t$, 
the phase velocity $v_\phi = \omega/k$, the group velocity $v_{\mathrm{g}} = \mathrm{d}\omega/\mathrm{d}k$, and $B_0/E_0$ are
\begin{align}
    & \dfrac{v_\phi}{c} = \dfrac{\omega}{ck} \approx 1 - \dfrac{1}{24}\left[1-\left(\dfrac{c\Delta t}{\Delta x}\right)^2\right]\left[\dfrac{\omega\Delta x}{c}\right]^2 \label{eqn:vphi} \\
    & \dfrac{v_{\mathrm{g}}}{c} = \dfrac{1}{c} \dd{\omega}{k} \approx 1 - \dfrac{1}{8}\left[1-\left(\dfrac{c\Delta t}{\Delta x}\right)^2\right]\left[\dfrac{\omega\Delta x}{c}\right]^2 \label{eqn:vg} \\
    & \dfrac{B_0}{E_0} \approx 1-\dfrac{1}{8}\left(\omega \Delta t\right)^2. \label{eqn:e_amp}
\end{align}
These expressions, with a time step given by $c\Delta t/\Delta x = 0.95/\sqrt{D}$ (to preserve numerical stability in $D$-dimensional simulations), will be used in the expressions for $\mathbf{E_1}$ and $\mathbf{B_1}$ employed in simulations.

The monochromatic plane-wave result can be extended to pulses by accounting for the frequency dependence of the parameters in Eqs.~\ref{eqn:vphi}-\ref{eqn:e_amp}. 
In the limit of small frequency bandwidth, the pulsed solution can be written as a slowly varying envelope multiplied by a rapidly varying phase, e.g.,
\begin{equation} \label{eqn:plane_func}
\begin{aligned} 
& E_y = \dfrac{1}{2}\Gamma\left(t-x/v_{\mathrm{g}}\right) e^{i\Phi\left(t-x/v_{\mathrm{g}}\right)} e^{ i\omega \left( t - x/v_\phi \right) }  + \mathrm{c.c.} = \Gamma \cos \left[ \Phi + \omega \left(t - x/v_\phi\right) \right]
\end{aligned}
\end{equation}
where $\Gamma$ and $\Phi$ represent the amplitude and phase of the pulse envelope. 
The small bandwidth limit assumes that $v_{\mathrm{g}}$ and $v_\phi$ can be evaluated solely at the central frequency of the pulse, which is valid for multicycle pulses under typically resolved simulation conditions (see Appendix~\ref{append:small}).

The magnetic field is related to the electric field by the numerical implementation of Faraday's Law and can be obtained from the relationship between the Fourier transformed field quantities $\mathbf{\Tilde{B}} = (ck/\omega) \nu \mathbf{\Tilde{E}}$, where $\nu \equiv (v_\phi/c)(B_0/E_0)$ is the numerical factor introduced by the field solver, with $\nu \to 1$ as $\Delta x \to 0$ and $\Delta t\to 0$. 
For an adequately resolved simulation, the dispersion introduced by $\nu$ can be neglected (i.e., $\nu$ can be evaluated at the central frequency of the laser pulse), in which case $\partial_t \mathbf{B} =-c\nu \nabla \times \mathbf{E}$. 
Assuming $\Gamma e^{i\Phi}$ is differentiable, the magnetic field can be written as the series solution to this equation,
\begin{align} \label{eqn:b_corr_full}
 B_z & = \dfrac{1}{2} \dfrac{B_0}{E_0} \Gamma e^{ i \Phi } e^{i\omega \left(t - x/v_\phi\right)} \left[1+\left(1-\dfrac{v_\phi}{v_{\mathrm{g}}}\right)\sum_{n=1}^\infty\left(\dfrac{i}{\omega}\right)^n\dfrac{1}{\Gamma e^{i\Phi}}\dd{^n\left(\Gamma e^{i\Phi} \right)}{t^n}\right] + \mathrm{c.c.}
 \end{align}
 In vacuum, it is typically the case that $v_\phi \approx v_{\mathrm{g}} + \mathcal{O}(\omega \Delta x/c)^n$ for an $n^\mathrm{th}$ order solver. 
 In the small bandwidth limit $[\mathrm{d}_t(\Gamma e^{i\Phi})]/(\Gamma e^{i\Phi}) \propto \Delta \omega \ll \omega$, 
 which makes the finite-duration correction terms in the summation significantly smaller than the numerical corrections to the phase and amplitude encoded in $v_\phi$, $v_{\mathrm{g}}$, and $B_0/E_0$.
 As a result,
 \begin{align} \label{eqn:plane_b}
 B_z \approx \dfrac{B_0}{E_0} \Gamma \cos \left[ \Phi + \omega \left( t - x/v_\phi \right) \right]
 \end{align}
 is usually adequate and has the advantage of being more computationally tractable than Eq.~\ref{eqn:b_corr_full}. 

\begin{figure}
    \centering
    \includegraphics[width=0.95\linewidth]{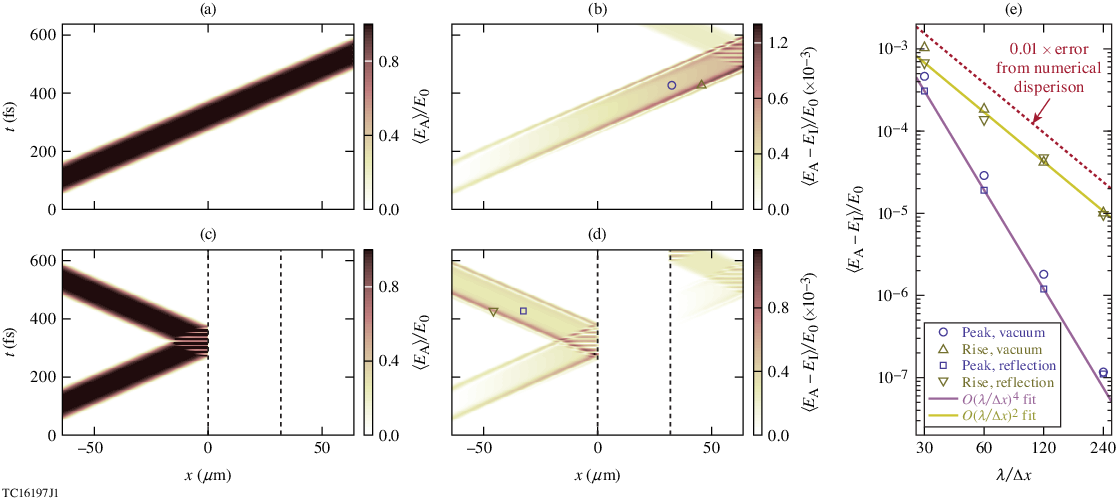}
    \caption{
    Demonstration of the analytic pulse technique (APT), using 1-D PIC simulations of a plane-wave laser pulse with a 100-fs FWHM, fourth-order super-Gaussian profile. (a)~Amplitude of the total electric field in vacuum with $\lambda/\Delta x = 30$, using APT. (b)~Difference between the total electric field using APT [$E_\mathrm{A}$, from (a)] and a simulation with the pulse launched from the boundary using the internal field update ($E_\mathrm{I}$). 
    (c)~Addition of a $n_\mathrm{e}=10$-$n_\mathrm{c}$ plasma slab (located between the black dotted lines) to the case from (a) for a pulse with a low (non-relativistic) amplitude.
    (d)~Difference between the total electric field in (c) and a simulation with the pulse launched from the boundary.
    (e)~Error in the total electric field, evaluated at the last time before the laser pulse hits the simulation boundary [markers in (b) and (d)].
    $\langle E (x) \rangle \equiv (2 \int_{x-\lambda/2}^{x+\lambda/2} \mathrm{d}x' E^2)^{1/2}$ is the moving cycle-average electric field amplitude. $n_\mathrm{c}\approx 10^{21}\;\mathrm{cm}^{-3}$ is the critical density for the wavelength $\lambda = 1\;\mu$m.
    }
    \label{fig:demo}
\end{figure}

The analytic pulse technique, using Eqs.~\ref{eqn:plane_func} and~\ref{eqn:plane_b} with $\Phi = 0$, was implemented in the open-source, MPI-parallel particle-in-cell code \textit{EPOCH}~\cite{arber2015epoch}. 
As shown in Fig.~\ref{fig:demo}, good agreement is achievable between simulations using APT and the traditional internal field update alone.
This agreement covers both laser propagation in vacuum [Figs.~\ref{fig:demo}(a) and~\ref{fig:demo}(b)] and laser--plasma interaction, even when the plasma response is significant [e.g., reflection of the pulse from overdense plasma, Figs.~\ref{fig:demo}(c) and~\ref{fig:demo}(d)].

In the limit that $\Delta x \to 0$ and $\Delta t \to 0$, the agreement between APT and traditional simulations is theoretically exact. For finite resolution, the implementation of APT presented here matches internally advanced pulses to the order of accuracy of the field solver at best. As shown in Fig.~\ref{fig:demo}(e), this choice, which is a matter of computational cost and convenience rather than a fundamental limitation of the approach, allows the differences between the analytic pulse and the internal field solver to remain more than two orders of magnitude lower than the error introduced by numerical dispersion. While the present formulation should be adequate for most use cases, it introduces a high-order mismatch between the phase velocity of the analytic pulse and that of the internal field solver, visible as a slowly growing difference in the relative phases as the laser pulses propagate in vacuum [e.g., in Fig.~\ref{fig:demo}(b)]. The phase slip is suppressed when the plasma is sufficiently dense for the phase velocity to be dominated by the plasma response rather than numerical dispersion [e.g., within the overdense plasma slab in Fig.~\ref{fig:demo}(d)], but restarts when the analytic pulse leaves the dense plasma region [e.g., to the right of the plasma slab in Fig.~\ref{fig:demo}(d)]. This may lead to increased resolution requirements when APT is used to model a significant distance of vacuum or very low density plasma on the rear side of an overdense target. Outside of this specific scenario, however, the physical fidelity of the simulation will be constrained by the order of accuracy of the field solver and cannot be improved upon with a more-accurate analytic phase velocity.

For the super-Gaussian pulse used in Fig.~\ref{fig:demo}, the error in the electric field between APT and simulations using a pulse launched from the simulation boundary was fourth order in $\omega \Delta x/c$ at the peak of the pulse and second order on its rising edge [Fig.~\ref{fig:demo}(e)]. The fourth-order scaling at the peak of the pulse reflects the aforementioned accuracy of the phase velocity, while the second-order scaling on its rising edge results from neglecting the bandwidth-related differences between APT and traditional simulations. 
Whereas traditional simulations used the incoming electromagnetic wave boundary condition, this boundary condition was only applied to the internally-advanced fields using APT. In \textit{EPOCH}, the incoming wave boundary condition has a slightly different dispersion than the bulk of the domain, and necessitated a second-order correction to the pulse amplitude to account for partial reflection of the pulse (see Appendix~\ref{append:boundary}). 
The dispersive nature of the boundary condition, however, was ignored. This, coupled with the omission of finite-duration and dispersive corrections in APT, results in second-order differences between the two approaches in the relatively high-bandwidth rising edge of the pulse.

While applying different boundary conditions to the analytic and internally-advanced fields is undesirable from the standpoint of exactly matching traditional simulations, it has the advantage of allowing APT to eliminate or reduce boundary-related errors.
For example, in Fig.~\ref{fig:demo}(b), unphysical reflection occurs at the $x_\mathrm{max}$ boundary in the traditional simulation, whereas this reflection is absent in the simulation using APT. The boundary condition can also be modified to operate on the total electric and magnetic fields rather than the internally-advanced fields alone if reproducing the same behavior as traditional simulations is desired. Such a modification would be useful for boundaries that lie beyond a reflective interface, for example to eliminate the unphysical reflection visible at the $x_\mathrm{max}$ boundary in Fig.~\ref{fig:demo}(d).

\begin{figure}
    \centering
    \includegraphics[width=0.8\linewidth]{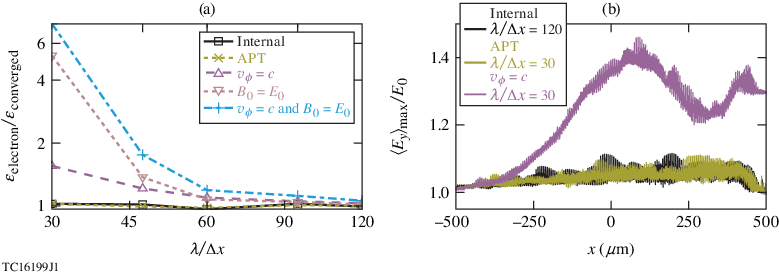}
    \caption{Importance of maintaining consistency between APT and the internal field solver. A relativistically intense ($a_0 = |e|E_0/mc\omega = 1$, where $m$ is the electron mass), $1$-$\mu$m-wavelength, 1-ps FWHM, fourth-order super-Gaussian plane-wave laser pulse interacts with a 1-mm linear density ramp (maximum electron density $n_\mathrm{e} = 0.01\;n_\mathrm{c}$; immobile ions), in 1D. (a)~Convergence of the average energy per electron, evaluated at the time when the total electron energy is greatest.  (b)~Maximum transverse electric-field amplitude at each point in the plasma, demonstrating substantial numerical Cerenkov in the case where $v_\phi = c$. Unless otherwise stated, $v_\phi$, $v_{\mathrm{g}}$, and $B_0$ are given by Eqs.~\ref{eqn:vphi} and~\ref{eqn:e_amp}. $n_\mathrm{c}\approx 10^{21}\;\mathrm{cm}^{-3}$ is the critical density for the laser wavelength. }
    \label{fig:wrong}
\end{figure}

As a final note, lowest-order numerical consistency between the analytic pulse and the internal field solver is crucial for maintaining the stability and accuracy of the PIC algorithm when particles are included. 
A naive attempt to include a laser pulse, for example, might use $v_\phi = c$ and $B_0 = E_0$, or add the second-order numerical corrections to only one of $v_\phi$ or $B_0/E_0$. 
The undesirable effects of failing to include numerical corrections in $v_\phi$ and $B_0/E_0$ are illustrated in Fig.~\ref{fig:wrong}. In this example, a long relativistic ($|e|E_0/mc\omega \gtrsim 1$) laser pulse undergoes self-modulation~\cite{antonsen1992smlwfa,sprangle1992smlwfa} during its propagation through a low-density plasma, driving electron heating in the combined fields of the pulse and the plasma response. The long time scale of the heating and the large distances involved render this configuration sensitive to compounding systematic errors. Unphysical heating results when the second-order corrections are neglected and produces slow convergence with increasing resolution [Fig.~\ref{fig:wrong}(a)]. In addition, numerical Cerenkov~\cite{godfrey1974cerenkov} is actively driven in the cases with $v_\phi = c$ [visible in $E_y$ in Fig.~\ref{fig:wrong}(b)], leading to the loss of energy conservation. When the second-order corrections are included, however, APT is in good agreement with traditional simulations.

\subsection{Performance} \label{sec:performance}

While evaluating analytic functions is more computationally expensive than addition and multiplication operations, simulations using the analytic pulse technique can have comparable cost to traditional simulations for identical domain size and resolution. The unique properties of APT can also provide flexibility-related cost savings by allowing the domain size or dimensionality to be reduced (discussed in Sec.~\ref{sec:applications}).

\begin{figure}
    \centering
    \includegraphics[width=0.8\linewidth]{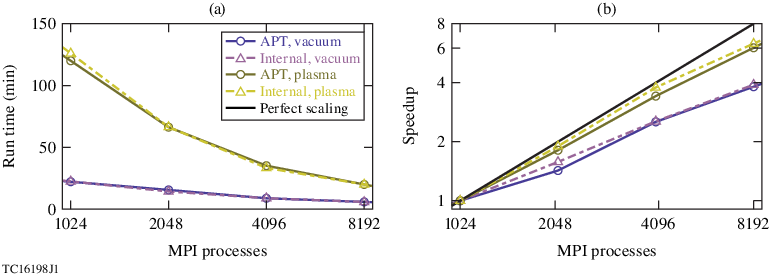}
    \caption{Strong scaling of simulations using APT compared to simulations with only the internal field update. (a)~Wall clock run time. (b)~Speedup relative to 1024 processes. The test simulation modeled a 3-D plane-wave laser pulse with peak amplitude $a_0 = 0.1$, 1-$\mu$m wavelength, and 100-fs temporal FWHM with $g=4$. The simulation domain was $64\times 8 \times 8\;\mu\mathrm{m}^3$ resolved by 30 cells per micron with $c \Delta t/\Delta x = 0.95/\sqrt{3}$, with open longitudinal and periodic transverse boundary conditions. The plasma, when present, consisted of a $56$-$\mu$m-long slab centered in the domain with electron density $n_\mathrm{e} = 10^{-6}\;n_\mathrm{c}$, and was represented by four electrons and four immobile ions per cell with second-order current deposition. $n_\mathrm{c}\approx 10^{21}\;\mathrm{cm}^{-3}$ is the critical density for the laser wavelength.}
    \label{fig:scaling}
\end{figure}

Figure~\ref{fig:scaling} shows the results of strong scaling studies conducted on the Intel KNL nodes of the Texas Advanced Computing Center supercomputer Stampede2 using 64 MPI processes per node. In the cases tested, simulations using APT had nearly identical run time [Fig.~\ref{fig:scaling}(a)] and scaling [Fig.~\ref{fig:scaling}(b)] to simulations using the internal field update alone.
The test problem consisted of a 3-D plane-wave laser pulse propagating either in vacuum or through a low-density plasma with a small number of particles per cell. The laser pulse was either launched from a simulated antenna at the domain boundary and advanced using the internal field update, or incorporated via evaluation of Eqs.~\ref{eqn:plane_func} and~\ref{eqn:plane_b} at every point in the domain at every time step. In the later case, the code was modified to add the analytically-specified fields to the internally advanced fields prior to the particle push. Details of the test simulation are given in the caption of Fig.~\ref{fig:scaling}.

For this test, and for all the other simulations in this work, the analytic expressions for $\mathbf{E_1}$ and $\mathbf{B_1}$ were coded in \textit{EPOCH} while allowing run-time specification of input parameters such as $E_0$, $\omega$, super-Gaussian order $g$, temporal duration $\tau$ (at which the electric field is reduced by $1/e^{2g}$), delay of the temporal envelope, and initial phase at a specified reference position. This configuration allowed for flexibility without sacrificing performance. Run-time specification of the functions for the fields themselves, while desirable from the standpoint of flexibility, was avoided to bypass \textit{EPOCH}'s function interpreter, which was found to otherwise dominate the simulation run time. A more-efficient parser and interpreter scheme, such as just-in-time compilation, may remove the requirement for hard coding.

Evaluating simple analytic functions, which does not require communication between processes, was found to be less expensive than the 3-D internal field update. In lower dimensionality, or with more-complex laser pulses, the function evaluation can become comparable in cost to the internal field update. In simulations with particles, however, the particle push typically dominates the simulation cost [e.g., the order-of-magnitude difference in run time between the cases with and without particles in Fig.~\ref{fig:scaling}(a)]. The analytic pulse technique is thereby not anticipated to significantly increase the expense of modeling laser--plasma interactions under most conditions (as is the case for all the simulations considered in Sec.~\ref{sec:applications}), and may instead offer flexibility-related cost savings.

\section{Example applications} \label{sec:applications}

The flexibility introduced by the analytic pulse technique enables simulations of otherwise prohibitive configurations and can introduce significant cost savings in established areas. 
The broad applicability of the technique can be seen in common laser--plasma scenarios covering a variety of regions in the parameter space of laser and plasma conditions.
First, two scenarios are considered where APT facilitates direct comparison between full and approximate solutions to Maxwell's equations, e.g., the paraxial approximation frequently employed in theory (Sec.~\ref{sec:paraxial}). 
Second, APT is shown to enable reduced-dimensionality simulation when the plasma dynamics, but not the laser pulse, are 1-D--like, which provides an orders-of-magnitude cost savings (Sec.~\ref{sec:dims}). 
Lastly, APT can significantly reduce the domain size needed to model laser pulses with spatiotemporal structure, enabling simulation of previously inaccessible configurations with desirable physical properties (Sec.~\ref{sec:ff}).

\subsection{Evaluating approximate solutions to Maxwell's equations} \label{sec:paraxial}

Theoretical work in laser--plasma interactions with non-planar laser pulses often utilizes approximate, analytically convenient solutions to Maxwell's equations. Meanwhile, the PIC simulations used to validate theoretical predictions are typically constrained to solve the full (discretized) set of Maxwell's equations. 
The ability of the analytic pulse technique to include approximate solutions for the fields of a laser pulse in an otherwise self-consistent simulation is therefore a valuable tool for verifying theoretical predictions.

As an example, consider a commonly employed paraxial approximation in which the laser pulse is assumed to be composed of a slowly varying, separable envelope modulated by a rapidly varying phase. For a monochromatic (temporally infinite) linearly polarized pulse, the vacuum wave equation can be written as
\begin{align}
    & \mathbf{E} = \dfrac{1}{2} \mathbf{F}\left(\mathbf{x}\right) e^{i\omega(t-x/v_\phi)} + \mathrm{c.c.} \\
    & \nabla_\perp^2 \mathbf{F} + \pd{^2}{x^2} \mathbf{F} - 2i \dfrac{\omega}{v_\phi} \pd{}{x}\mathbf{F} = 0
\end{align}
where $\nabla^2_\perp = \partial_y^2+\partial_z^2$ and $v_\phi$ accounts for the effect of discretization. In the paraxial approximation, $|\partial_x \mathbf{F}| \ll |\omega \mathbf{F}/v_\phi|$, and the wave equation reduces to
\begin{align} \label{eqn:paraxial_diff}
    & \nabla_\perp^2 \mathbf{F} - 2i\dfrac{\omega}{v_\phi} \pd{}{x}\mathbf{F} \approx 0. 
\end{align}

Consider a linearly polarized laser pulse in which Eq.~\ref{eqn:paraxial_diff} applies to the primary electric-field component $E_y$. 
Laguerre--Gaussian modes are a commonly employed set of solutions to the paraxial wave equation, with the lowest-order (Gaussian) mode given by
\begin{align}
    & F = F_0 \left(\dfrac{w_0}{w}\right)^{\frac{D-1}{2}} \exp\left(-\dfrac{r^2}{w^2}\right) \exp\left[i\dfrac{kr^2}{2R} - \dfrac{i\left(D-1\right)}{2}\mathrm{atan}\left(\dfrac{x_f-x}{x_\mathrm{R}}\right)\right], \label{eqn:paraxial} \\
    & w = w_0 \sqrt{1+\left(x_f-x\right)^2/x_\mathrm{R}^2}, \\
    & R = \left(x_f-x\right)\left[1+\dfrac{x_\mathrm{R}^2}{\left(x_f-x\right)^2}\right], 
\end{align}
where $D-1$ is the number of transverse dimensions, $x_f$ is the focal location, and $x_\mathrm{R} = k w_0^2/2$ is the Rayleigh range.

The monochromatic result can then be extended to pulsed solutions via Fourier transform. When the frequency dependence of $x_\mathrm{R}$ can be ignored, $F$ may be excluded from the Fourier transform integral. In the small frequency bandwidth limit, the primary component of the electric field is then simply $F$ multiplied by the plane-wave expression (prior to the addition of the complex conjugate in Eq.~\ref{eqn:plane_func}). This is the usual separable paraxial limit and requires $|x-x_f| \ll (\omega/\Delta \omega)(v_{\mathrm{g}}/ v_\phi) |F/\partial_x F| \sim \omega \tau x_\mathrm{R}  v_{\mathrm{g}}/v_\phi$, where $\tau$ is the transform--limited pulse duration.

The secondary field component $E_x$ is then approximately given by $\nabla \cdot \mathbf{E} \approx -i \omega E_x/v_\phi + \partial_y E_y = 0$.
Meanwhile, the magnetic field is determined by Faraday's Law, the $z$ component of which is $\partial_t B_z = - c\nu_x \partial_x E_y + c\nu_y \partial_y E_x$, where $\nu$ is determined by the resolution in each direction.
The paraxial approximation requires $k w_0 \gg 1$, meaning $|\partial_y E_x| \ll |\partial_x E_y|$ and this term may be dropped in the interest of simplifying the solution. 
In this limit, $B_z$ is given by $F$ multiplied by the plane-wave result (prior to the addition of the complex conjugate in Eq.~\ref{eqn:plane_b}), while $B_x$ is found from $\nabla \cdot \mathbf{B} \approx -i \omega B_x/c + \partial_z B_z = 0$.

\begin{figure}
    \centering
    \includegraphics[width=0.95\linewidth]{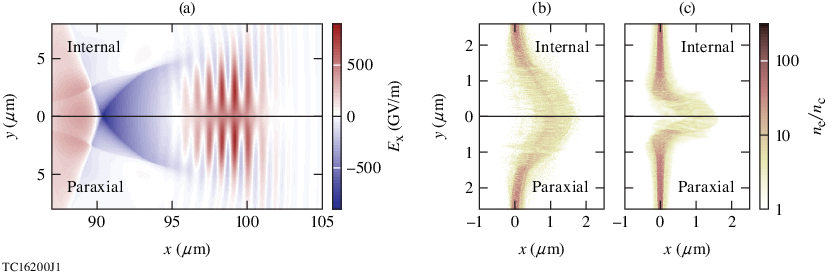}
    \caption{Sensitivity of plasma dynamics to the paraxial approximation. Top: Full solutions to Maxwell's equations using the internal field update with incoming wave boundary conditions. Bottom: Paraxial analytic pulse technique. (a)~Wakefield regime, demonstrating good agreement for a 15-fs and 8-$\mu$m FWHM Gaussian pulse with $a_0 = 3$ in underdense plasma with electron density $n_\mathrm{e} = 0.01 \; n_\mathrm{c}$. [(b),(c)]~Radiation pressure acceleration, demonstrating good agreement at (relatively) large spot size and modest differences at smaller spot size, for a 15-fs Gaussian pulse with $a_0 = 20$ interacting with a 50-nm-thick plastic target (fully ionized CH, $n_\mathrm{e} = 180\;n_\mathrm{c}$) with a FWHM laser spot size of (b) 3~$\mu$m or (c) 1~$\mu$m. $n_\mathrm{c}\approx 10^{21}\;\mathrm{cm}^{-3}$ is the critical density for the laser wavelength. }
    \label{fig:paraxial}
\end{figure}

The assumptions underlying the separable paraxial approximation cause the analytic pulse described above to depart from the full solution to Maxwell's equations produced by the internal field update. 
For example, the paraxial assumption $|\partial_x F| \ll |\omega F/v_\phi|$
is not valid under tight focusing conditions, which is particularly relevant for applications requiring ultrahigh intensity, such as radiation pressure acceleration~\cite{bulanov2016rpa}, inverse Compton scattering~\cite{harvey2016tight}, and pair-plasma generation~\cite{jirka2017pair,tamburini2017pair}. 
Simulations with tight focusing are additionally complicated by the breakdown of the normal-incidence boundary or antenna conditions commonly used to launch pulses, which if unmodified lead to numerical shortening of the focal length and asymmetry about the focus~\cite{thiele2016tight}.
Departure from the paraxial approximation and boundary-condition--related error, however, does not automatically guarantee substantive changes to the dynamics of laser--plasma interaction, and should be investigated to evaluate the validity of theory based on paraxial pulses.
Comparison between paraxial beams, beams launched using the ordinary boundary conditions, and beams launched with modified, closer-to-paraxial boundary conditions~\cite{thiele2016tight} is uniquely enabled by the analytic pulse technique.

To demonstrate this new capability,
the paraxial solutions for $E_x$, $E_y$, $B_x$, and $B_z$ discussed above were implemented in APT. Comparisons were then made between simulations using paraxial APT and simulations with pulses launched from the simulation boundary using the standard normal-incidence boundary condition. 
Parameters were chosen to probe the sensitivity of the plasma dynamics to the paraxial approximation for two common laser--plasma applications: laser wakefield acceleration and radiation pressure acceleration.

The paraxial approximation is expected to hold for sufficiently large spot sizes, for example in laser wakefield acceleration [Fig.~\ref{fig:paraxial}(a)] and the larger-spot radiation pressure acceleration case [Fig.~\ref{fig:paraxial}(b)]. In these cases, good agreement was observed between APT and all-internal simulations. 
In the smaller-spot radiation pressure acceleration case [Fig.~\ref{fig:paraxial}(c)], discrepancies between the paraxial pulse and the all-internal pulse had a modest impact on the plasma dynamics, leading to a somewhat larger transverse extent of the deformation in the all-internal simulation. 

While a comprehensive study of paraxial effects in simulations with small laser spot size is left for future work, Figs.~\ref{fig:paraxial}(a) and~\ref{fig:paraxial}(b) demonstrate that, as one would expect,
paraxial APT can safely be employed when the spot size is sufficiently large.
The ability of APT to produce an identical physics result to traditional simulations can then be used to significantly reduce simulation cost when the resource needs are predominantly determined by the laser pulse rather than the plasma.
Examples of how the flexibility introduced by APT can lead to cost savings are given in Secs.~\ref{sec:dims} and~\ref{sec:ff}.

\subsection{Reduced-dimensionality simulation} \label{sec:dims}

While the focusing of a laser pulse is always multidimensional, the particle motion driven by a focusing pulse can sometimes be modeled in fewer spatial dimensions than the pulse itself. 
For example, the trajectories of individual electrons in the inverse Compton scattering~\cite{sokolov2010compton,bulanov2013compton} of a sufficiently high-energy electron beam from a laser pulse can be effectively 1D-3V (one spatial and three velocity dimensions), but the radiation they emit depends upon the focusing-related variation of the laser intensity and would not be properly modeled in a traditional 1-D simulation~\cite{blackburn2017pair}. 
One-dimensional models have also been proposed to treat Breit-Wheeler pair production from inverse Compton-generated photons when the angular photon spectrum is narrow, a process that is highly sensitive to laser intensity~\cite{blackburn2017pair,amaro2021pair}. 
However, the need for full self-consistency in the usual internal electromagnetic field update means that 3-D or quasi-3-D simulations are required to correctly model the intensity profile of a focusing laser pulse, which imposes high computational cost for these applications. 

The analytic pulse technique can incorporate lineouts or slices of laser pulses undergoing 2-D or 3-D focusing in lower-dimensional simulations, resulting in significant computational savings. For example, consider a series of 1D-3V simulations representing different locations in a plane transverse to the propagation direction of a laser pulse. The plasma dynamics driven by this pulse will be captured properly using APT in 1D if particle displacement in the transverse directions is not important and if the electric and magnetic fields remain consistent with the 3-D case. The latter condition holds if transverse gradients in the plasma density and current (e.g., as would contribute to self-focusing) have a negligible effect on the fields of the driving laser pulse.

\begin{figure}
    \centering
    \includegraphics[width=0.95\linewidth]{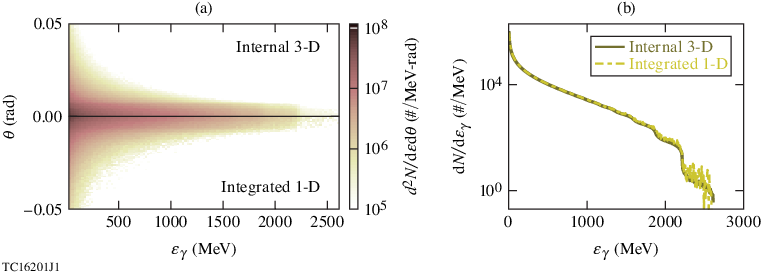}
    \caption{Reduced-dimensionality simulation using APT, for the inverse Compton scattering of a 2.6-GeV electron beam counter-propagating with a laser pulse with $a_0=50$. Both the laser and electron density profile were Gaussian with a 15-fs duration and a 3-$\mu$m-FWHM spot size. (a)~Angularly resolved photon energy spectrum. $\theta$ is the polar angle, relative to electron propagation direction. (b)~Angularly integrated photon energy spectrum. 1-D results were obtained by numerically integrating the spectra from 30 simulations equally spaced in $r$ with $\Delta r = 0.1\;\mu$m and ten times the number of particles per cell used in 3D.}
    \label{fig:dim}
\end{figure}

The effectiveness of reduced-dimensionality simulations is demonstrated in Fig.~\ref{fig:dim}. 
Photon generation by the inverse Compton scattering of a high energy electron beam from an intense laser pulse was considered both in a full 3-D simulation and in a series of 1-D simulations sampling different radii in the laser pulse. In 1D, analytic paraxial pulses were employed, using both the primary ($E_y$, $B_z$) and secondary ($E_x$, $B_x$) field components (see Sec.~\ref{sec:paraxial}). The resulting photon spectra were then integrated radially for comparison with the 3-D results. As expected, the 1-D--like nature of the interaction led to excellent agreement in the photon spectra between the 3-D simulation and the series of 1-D simulations using APT [Fig.~\ref{fig:dim}].

Reducing the simulation dimensionality, when possible, is highly desirable. Even with equal resolution, a series of 1-D simulations will be more efficient than its multi-dimensional counterpart due to reduced communication and related overhead. Moreover, simulations using APT only need to sample the particle statistics and can be conducted over a smaller transverse domain and with nonuniform spacing, if desired.
Whereas the 3-D simulation in Fig.~\ref{fig:dim} required several hours on hundreds of nodes on a supercomputer, the series of 1-D simulations was completed in less than one minute on a desktop, resulting in an approximately 40,000-fold savings. Such significant savings is uniquely enabled by the analytic pulse technique and is expected to facilitate parameter scans and prototyping that would otherwise be prohibitively expensive. Inexpensive accurate or semi-accurate simulations are also attractive for machine learning including multi-fidelity transfer learning and Bayesian optimization~\cite{djordjevic2023ml,irshad2023ml}.

\subsection{Spatiotemporally shaped laser pulses} ~\label{sec:ff}

The analytic pulse technique bypasses the limitations of the methods traditionally used to introduce laser pulses into the simulation domain and can facilitate the modeling of novel pulse types.
Spatiotemporal coupling, for instance, uses correlation between the spatial and temporal degrees of freedom of a laser pulse to control the intensity dynamics separately from the laser phase and group velocities~\cite{kondakci2017spatio,sainte-marie2017spatio,froula2018ff,palastro2020dephasingless}. This allows the evolution of the laser pulse to be optimized for, rather than determined by, underlying physical processes.
For example, a recently demonstrated form of pulses with spatiotemporal coupling with a moving focal position, known as flying-focus pulses, have been proposed to push laser wakefield acceleration beyond the dephasing limit~\cite{palastro2020dephasingless}. These pulses can also strongly enhance nonlinear Thomson scattering by countering ponderomotive deceleration~\cite{ramsey2022nlts}.
However, the very property that makes flying-focus pulses desirable for applications, i.e., the moving focal location, also makes them challenging to simulate using traditional approaches.

\begin{figure}
    \centering
    \includegraphics[width=0.95\linewidth]{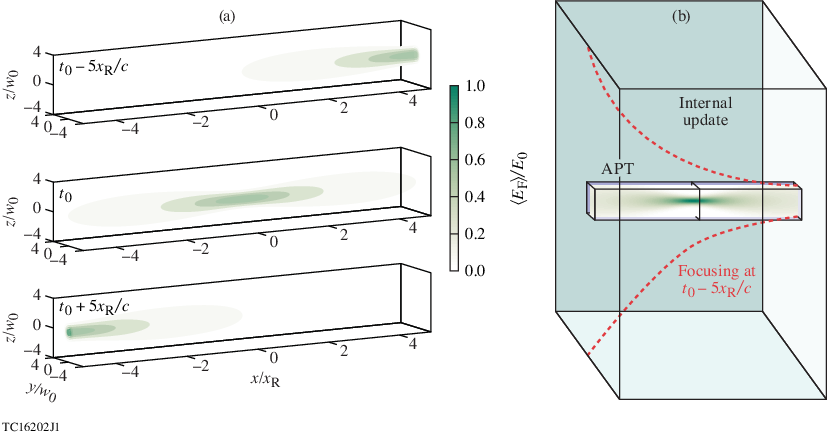}
    \caption{Implementation of flying-focus pulses. (a)~Cycle-averaged field amplitude for a 3-$\mu$m and 1-ps FWHM backward-moving Gaussian flying focus with $\beta_f = -1$ and $\lambda = 1\;\mu$m, capturing the motion of the focal spot over ten Rayleigh ranges. (b)~Simulation domain required to capture the field in (a) when the pulse is (green) launched from the simulation boundary and the box must capture the focal dynamics over the full propagation distance or (blue) incorporated with APT and the box need only capture the intense part of the pulse.}
    \label{fig:ff}
\end{figure}

Similar to the stationary focusing discussed in Sec.~\ref{sec:paraxial}, the wave equation admits non-stationary focusing solutions of the form
\begin{align}
    & \mathbf{E} = \dfrac{1}{2} \mathbf{F}\left(\xi,y,z\right) e^{i\omega(t-x/v_\phi)} + \mathrm{c.c.} \\
    & \nabla_\perp^2 \mathbf{F} + \left( 1-\beta_f^2\right) \pd{^2}{\xi^2} \mathbf{F} - 2i\dfrac{\omega}{c}\left(1-\beta_f\right) \pd{}{\xi}\mathbf{F}= 0,
\end{align}
where $\xi = x - v_f t$, with (constant) intensity peak velocity $v_f$, and $\beta_f = v_f/c$.
With the assumption $|(1+\beta_f) \partial_\xi \mathbf{F}| \ll |\omega  \mathbf{F}/c|$, the above can be reduced to a modified monochromatic paraxial wave equation~\cite{ramsey2022nlts},
\begin{align}
    & \nabla_\perp^2 \mathbf{F} - 2i\dfrac{\omega}{c}\left(1-\beta_f\right) \pd{}{\xi}\mathbf{F} = 0,
\end{align}
which admits similar solutions to the ordinary paraxial wave equation, including the flying-focus equivalent of Laguerre--Gaussian solutions.

To illustrate the difficulty of simulating such pulses, consider, for example, a monochromatic Gaussian flying-focus solution in the modified paraxial limit, which is given by Eq.~\ref{eqn:paraxial} with the transformation $x_\mathrm{R} \to x_\mathrm{R}' = (1-\beta_f) x_\mathrm{R}$ and $x_f-x \to \xi_0 - \xi$, where $\xi_0 = x_0 -v_f t_0$ for a focus at $x_0$ at $t_0$. As in the stationary case, pulsed solutions can be constructed via Fourier transform. In the separable limit, the frequency dependence of $x_\mathrm{R}'$ is neglected, encoding the assumption $|\xi - \xi_0 | \ll (\omega/\Delta \omega)(v_{\mathrm{g}}/v_\phi)|F/\partial_\xi F| $. The spatial width of the intensity peak at any given time is then approximately $|x_\mathrm{R}'|$, which is the flying-focus analog of the Rayleigh range. An example of such a pulse with $\beta_f = -1$ is shown in Fig.~\ref{fig:ff}(a).

Laguerre--Gaussian flying-focus pulses must propagate over a distance larger than $|x_\mathrm{R}'|$ for the effect of the moving focus to be observable. Consider a case in which the propagation of a flying-focus pulse over a distance $L \gg |x_\mathrm{R}'|$ must be captured in the simulation domain.
For forward-moving pulses with $\beta_f \sim 1$, $|x_\mathrm{R}'|$ is significantly smaller than $x_\mathrm{R}$, and the simulation domain may not need to cover multiple Rayleigh ranges. Such pulses may be introduced using either boundary or initial conditions (e.g., through linear superposition~\cite{pierce2023astrl}).
In the more-general case, however, simulations will need to capture $L \gg x_\mathrm{R}$. If the pulse is either launched from a single simulation boundary or initialized in its entirety within the domain, the domain must be sufficiently large to capture the effective beam waist at a distance $L$ from the focus, i.e., the required radius of the domain is $R_\mathrm{I} \sim w_0 \sqrt{1+L^2/x_\mathrm{R}^2} \sim w_0 L/x_\mathrm{R}$, leading to a very large simulation box [e.g., the green box in Fig.~\ref{fig:ff}(b)]. 
Multiboundary antennae have been developed to sidestep this issue for a fixed focal position using the total field/scattered field technique and plane-wave decomposition~\cite{merewether1980tfsf,schneider2006tfsf,tan2010tfsf,capoglu2013tfsf}. Extending these antennae to pulses with spatiotemporal coupling, however, is nontrivial and may also be prohibitively expensive due to the enhanced, spatially and temporally varying spectral content of the pulse.

If instead the pulse is incorporated using the analytic pulse technique, the simulation domain need only capture the high-intensity part of the pulse, with radius $R_\mathrm{A} \sim w_0 \ll R_\mathrm{I}$. As illustrated in Fig.~\ref{fig:ff}(b), the reduction in the size of the domain (a factor of $L^2/x_\mathrm{R}^2$) is significant and results in 100-fold savings when $L/x_\mathrm{R} = 10$. This cost savings is anticipated to enable multidimensional simulations of flying-focus pulses under conditions where introduction of the pulse in the simulation domain using traditional approaches is prohibitively expensive. Similar to Secs.~\ref{sec:paraxial} and~\ref{sec:dims}, APT also enables direct evaluation of the approximations used to obtain an analytic form for flying-focus pulses and permits flying-focus pulses to be modeled in 1-D simulations when the plasma response is 1-D--like. Additionally, analytic flying-focus pulses can be modeled with a moving window following the intensity peak velocity rather than the phase velocity, which is especially advantageous for backward-moving flying-focus pulses. 
In this case, electromagnetic waves generally both enter and exit the domain, which may be completely filled with plasma. The implementation of this configuration can be significantly simplified with the addition of an analytic current.

\section{Analytic pulse technique with analytic current} \label{sec:j1}

As noted in Sec.~\ref{sec:applications}, laser--plasma applications frequently involve laser pulses focusing into and interacting with large regions of low-density plasma.
From a physics standpoint, kinetic modeling may only be required in some parts of the plasma, e.g., where the plasma response is nonlinear, while the remainder may be well treated by linear analytic models. At the same time, simulations are constrained by the need to capture focusing and to model the simulation domain as embedded in a larger plasma-filled volume. For traditional simulations, satisfying these constraints requires either simulating the entire extent of the laser pulse or using plasma-embedded boundary conditions~\cite{berenger1994pml,merewether1980tfsf,schneider2006tfsf,tan2010tfsf,capoglu2013tfsf}. The former approach is typically only feasible for short laser pulses with relatively small spot sizes, while the latter typically adds significant computational complexity and is nontrivial to extend to new laser pulse profiles (see Sec.~\ref{sec:ff}).
As an alternative, when the current on the simulation boundary is analytically calculable, e.g., for the linear plasma response characteristic of low-intensity laser--plasma interactions, Maxwell's equations may be split in a way that removes this current from the internal field solver. The analytic pulse technique with analytic current can then enable the use of vacuum boundary conditions for the internally-advanced fields, significantly simplifying the treatment of laser pulses in low-density plasmas.

With the addition of an analytic current $\mathbf{J_1}$, Eqs.~\ref{eqn:E_1} and~\ref{eqn:E_2} become
\begin{align}
    & \dfrac{1}{c}\pd{\mathbf{B_1}}{t} = - \nabla \times \mathbf{E_1} \;\; ; \;\; \dfrac{1}{c}\pd{\mathbf{E_1}}{t} = \nabla \times \mathbf{B_1}  + \dfrac{4\pi}{c}\mathbf{J_1} \label{eqn:E_1_j}\\
    & \dfrac{1}{c}\pd{\mathbf{B_2}}{t} = - \nabla \times \mathbf{E_2} \;\; ; \;\; \dfrac{1}{c}\pd{\mathbf{E_2}}{t} = \nabla \times \mathbf{B_2} + \dfrac{4\pi}{c} \left(\mathbf{J} - \mathbf{J_1} \right), \label{eqn:E_2_j} 
\end{align}
where $\mathbf{J}$ (no subscript) is still the physical current carried by the plasma.
Consider, for example, the linear plasma response to a monochromatic electromagnetic wave in which the current and fields are expressed in the form $\mathbf{J_1} = \frac{1}{2} \mathbf{J_{1,\mathrm{c}}}e^{i\omega(t-x/v_\phi)} + \mathrm{c.c.}$ The current is given by $4 \pi \mathbf{J_{1,\mathrm{c}}} = -i \omega_p^2 \zeta \mathbf{E_{1,\mathrm{c}}}/\omega$, where $\omega_p^2 = \Sigma_{j} 4\pi q_j^2 n_j/m_j$ is the sum of the squares of the plasma frequency for each species $j$ with charge $q_j$, density $n_j$, and mass $m_j$. 
The factor $\zeta(\omega,k)$ is introduced to account for the dependence of the phase velocity on the field gather and current deposition routines, with $\zeta \to 1$ in the limit that $\Delta x \to 0$ and $\Delta t \to 0$.
In the second-order explicit Yee scheme, with $\mathbf{J_1}$ evaluated at the half time step, the dispersion relation for Eqs.~\ref{eqn:E_1_j} is given by
\begin{align} \label{eqn:dispersion_j}
    \sin^2\left(\dfrac{\omega \Delta t}{2}\right) - \dfrac{\omega_p^2}{\omega^2} \left(\dfrac{\omega \Delta t}{2}\right) \sin \left(\dfrac{\omega \Delta t}{2}\right) \zeta \left(\omega,k\right) = \dfrac{c^2\Delta t^2}{\Delta x^2}\sin^2\left(\dfrac{k\Delta x}{2}\right)
\end{align}
and
\begin{align} \label{eqn:e_amp_j}
    B_0 = \cos\left(\omega \Delta t/2\right) \sqrt{1 - \dfrac{\omega_p^2}{\omega^2} \dfrac{\zeta \omega \Delta t/2 }{\sin \left(\omega \Delta t/2\right)}} E_0.
\end{align}

As in the vacuum case, an approximate pulsed solution to Maxwell's equations can be constructed via Fourier transform, resulting in a similar expression to Eq.~\ref{eqn:plane_func} in the small frequency bandwidth limit. Consider an electric field 
\begin{align} \label{eqn:plane_e_j}
& E_y = \Gamma\left(t-x/v_{\mathrm{g}}\right) \cos \left[ \Phi\left(t-x/v_{\mathrm{g}}\right) + \omega\left( t - x/v_\phi \right) \right].
\end{align}
The current is given by $\mathrm{d}\mathbf{J}/\mathrm{d}t = \omega_p^2 \zeta \mathbf{E}/4 \pi$, resulting in
\begin{align}
 J_y & = -i \dfrac{\omega_p^2 \zeta}{4 \pi \omega}  \Gamma e^{ i \Phi} e^{ i\omega \left(t - x/v_\phi\right) } \left[1+\sum_{n=1}^\infty\left(\dfrac{i}{\omega}\right)^n\dfrac{1}{\Gamma e^{i\Phi}}\dd{^n\left(\Gamma e^{i\Phi} \right)}{t^n}\right] + \mathrm{c.c.} \\
 & \begin{aligned}
     \approx \dfrac{\omega_p^2 \zeta}{4 \pi \omega} & \left[ 1- \dfrac{1}{\omega^2 \Gamma}\dd{^2\Gamma}{t^2}+\dfrac{1}{2}\left(\dfrac{1}{\omega \Gamma}\dd{\Gamma}{t}\right)^2 - \dfrac{1}{\omega} \dd{\Phi}{t} + \dfrac{1}{\omega^2}\left(\dd{\Phi}{t}\right)^2  \right] \Gamma \\
      & \times \sin \left[ \Phi +  \omega \left(t - x/v_\phi\right) + \dfrac{1}{\omega \Gamma}\dd{\Gamma}{t} - \dfrac{1}{\omega^2} \dd{^2\Phi}{t^2} - \dfrac{2}{\omega^2} \left(\dfrac{1}{\Gamma}\dd{\Gamma}{t}\right) \left( \dd{\Phi}{t} \right) \right],
 \end{aligned} \label{eqn:j1}
\end{align}
where the second expression includes only the lowest-order finite-duration corrections in amplitude and phase, 
and makes use of $\exp(ix)\approx 1+ix-x^2/2$. Similarly, the magnetic field is related to the electric field by the numerical implementation of Faraday's Law. 
In the analytic current case, $v_\phi \neq v_{\mathrm{g}}$ due to physical as well as numerical dispersion. The finite-duration correction is only negligible in very underdense plasma, where vacuum-like dispersion is recovered. Otherwise, at least the lowest order terms in the sum in Eq.~\ref{eqn:b_corr_full} should be kept.
Retaining only the lowest-order corrections to the amplitude and phase yields
\begin{align} \label{eqn:b_corr_t}
& \begin{aligned}
     B_z \approx \dfrac{B_0}{E_0} & \left[ 1 - \dfrac{\left(1-v_\phi/v_{\mathrm{g}}\right)}{\omega^2 \Gamma}\dd{^2\Gamma}{t^2}+\dfrac{1}{2}\left(\dfrac{\left(1-v_\phi/v_{\mathrm{g}}\right)}{\omega \Gamma}\dd{\Gamma}{t}\right)^2 - \dfrac{\left(1-v_\phi/v_{\mathrm{g}}\right)}{\omega} \dd{\Phi}{t} \right. \\
     & \left. \phantom{1-} + \dfrac{\left(1-v_\phi/v_{\mathrm{g}}\right)}{\omega^2}\left(\dd{\Phi}{t}\right)^2  \right] \Gamma  \\
     & \times \cos \left[ \Phi + \omega \left(t - x/v_\phi\right) + \dfrac{\left(1-v_\phi/v_{\mathrm{g}}\right)}{\omega \Gamma}\dd{\Gamma}{t} - \dfrac{\left(1-v_\phi/v_{\mathrm{g}}\right)}{\omega^2} \dd{^2\Phi}{t^2} \right. \\
     & \left. \phantom{\times \cos \Phi + } - \dfrac{2\left(1-v_\phi/v_{\mathrm{g}}\right)}{\omega^2} \left(\dfrac{1}{\Gamma}\dd{\Gamma}{t}\right) \left( \dd{\Phi}{t} \right) \right].
\end{aligned}
\end{align}

The small frequency bandwidth limit breaks down when group velocity dispersion (GVD) becomes important. The breakdown of this limit requires both violation of the monochromatic approximation made in deriving Eq.~\ref{eqn:e_amp_j} (see Appendix~\ref{append:dispersion}) and sensitivity to GVD-related pulse evolution. The latter condition is not well suited to the second-order explicit Yee scheme, in which GVD only dominates over the numerical effects of imperfect dispersion at high resolution (see Appendix~\ref{append:dispersion}). Simulations using the second-order explicit Yee scheme, such as those included in this section, are often adequately resolved in terms of physical observables at significantly lower resolution, where numerical effects are dominant. Under such conditions, GVD corrections are not appropriate.

\begin{figure}
    \centering
    \includegraphics[width=0.95\linewidth]{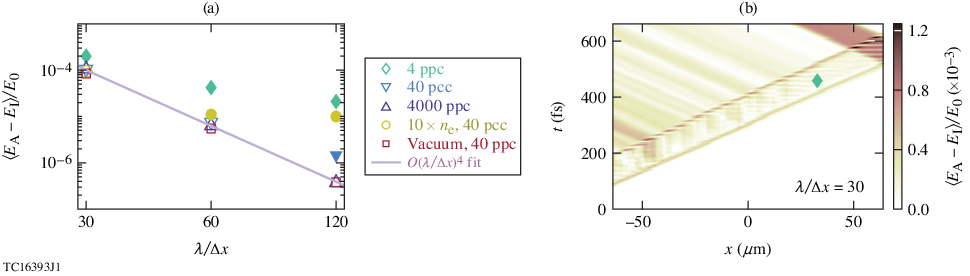}
    \caption{Demonstration of APT with analytic current. (a)~Difference in the total electric fields between APT ($E_\mathrm{A}$) and traditional simulations ($E_\mathrm{I}$), using the same laser pulse as Fig.~\ref{fig:demo}, in the low-intensity (non-relativistic) regime. Filled symbols represent cases that are dominated by statistical particle noise. The domain is filled by an $n_\mathrm{e} = 10^{-3}\,n_\mathrm{c}$-plasma unless otherwise specified. The cases labeled ``vacuum'' consist of propagation in vacuum following a linear ramp of $6\,c/\omega_p$ down from density $n_\mathrm{e} = 10^{-2}$ (used for $\mathbf{J_1}$ in APT, see text). (b)~Example of the difference between the total fields from the 4 ppc, $\lambda/\Delta x = 30$ cases in~(a), showing the effect of statistical particle noise.
    $\langle E (x) \rangle \equiv (2 \int_{x-\lambda/2}^{x+\lambda/2} \mathrm{d}x' E^2)^{1/2}$ is the moving cycle-average electric field amplitude. $n_\mathrm{c}\approx 10^{21}\;\mathrm{cm}^{-3}$ is the critical density for the laser wavelength.
    }
    \label{fig:j2_demo}
\end{figure}

As shown in Fig.~\ref{fig:j2_demo}, APT with analytic current can provide good agreement with traditional simulations. For this comparison, two physical configurations were considered. First, propagation of the laser pulse was modeled through uniform plasma. In APT, the laser pulse was launched directly into the plasma-filled domain by setting the density used in $\mathbf{J_1}$ equal to the physical plasma density. In traditional simulations, the pulse was launched using the incoming wave boundary condition into several wavelengths worth of vacuum before encountering a sharp jump in density up to the value in the remainder of the domain. Second, the laser pulses were modeled with a short density down ramp followed by propagation in vacuum to demonstrate that APT can match traditional simulations even if the density changes significantly within the simulation domain. For APT, the density down-ramp was placed at the simulation boundary and $\mathbf{J_1}$ used the density value at the boundary. The same density ramp was added to the traditional simulations, which now consisted of a few wavelengths of vacuum, a sharp density jump up and short ramp down, and then vacuum for the remainder of the domain. While the initial amplitude of the internally-advanced pulse was modified in both cases to account for the physical reflectivity of the sharp vacuum--plasma interface and the effect of the boundary condition, the shape of the pulse envelope was not corrected. The resulting pulse envelope represents a slightly different---although equally valid---solution to Maxwell's equations than the analytic case (see Appendix~\ref{append:boundary}), for which reason the comparisons in Fig.~\ref{fig:j2_demo} were only made near the peak of the super-Gaussian pulse. 

\begin{figure}
    \centering
    \includegraphics[width=0.95\linewidth]{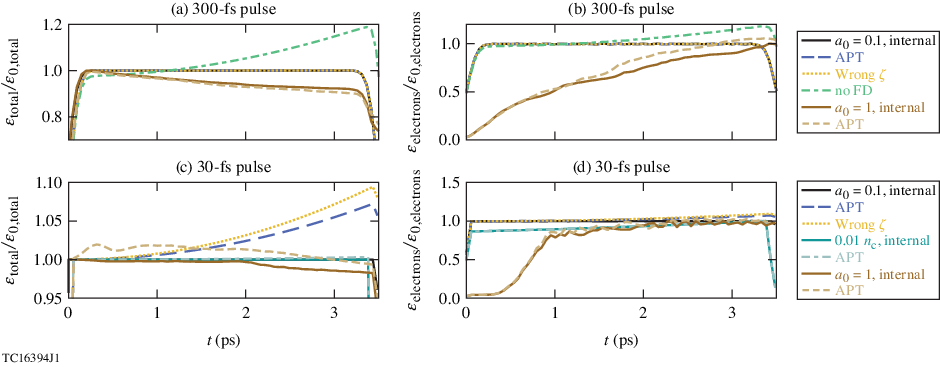}
    \caption{Energy conservation in APT with analytic current. In the nominal case, a temporally Gaussian plane-wave laser pulse with a 300-fs [(a) and (b)] or 30-fs [(c) and (d)] duration propagates through 1024~$\mu$m of plasma with $n_\mathrm{e} = 0.05\,n_\mathrm{c}$ and fixed ions. [(a),(c)] Total energy in the simulation domain. [(b),(d)] total energy in electrons. Energies are normalized to the maximum value from the corresponding traditional simulation (labeled ``internal''). Simulations are grouped in the legend by the peak $a_0$. The cases labeled ``wrong $\zeta$'' use a different value of $\zeta$ than the internal field solver and the case labeled ``no FD'' omits the finite-duration correction from Eqs.~\ref{eqn:j1} and~\ref{eqn:b_corr_t}.}
    \label{fig:j2_energy}
\end{figure}

Unlike in the vacuum formulation of APT, observation of the fourth-order scaling of the numerical error at the peak of the pulse is restricted to a very low density and a very large number of particles per cell [open symbol cases in Fig.~\ref{fig:j2_demo}(a)]. Outside of these conditions, it is not possible to obtain exact agreement in the fields on a cell-by-cell basis due to the effect of particle statistics [e.g., the fluctuations visible in Fig.~\ref{fig:j2_demo}(b)]. Nevertheless, APT with analytic current produces statistically correct plasma dynamics when the pulse is sufficiently well modeled, as shown in Fig.~\ref{fig:j2_energy}. 

When the analytic pulse and current constitute an accurate solution to Maxwell's equations, the analytic pulse technique has robust energy conservation [Fig.~\ref{fig:j2_energy}(a)] and converges similarly to traditional simulations [e.g., as measured by electron energization, Fig.~\ref{fig:j2_energy}(b)]. These properties hold even if the effects of the field gather and current deposition routines on the
phase velocity are not correctly accounted for [case labeled wrong $\zeta$ in Figs.~\ref{fig:j2_energy}(a) and~\ref{fig:j2_energy}(b)]. 
Agreement between APT and traditional simulations also requires the vacuum boundary condition to be appropriate, which for the linear plasma response given by Eq.~\ref{eqn:j1} requires non-relativistic pulses ($a_0 = |e|E_0/mc\omega \lesssim 0.1$).
Energy is still well conserved for higher-intensity pulses, but the incoming pulse is affected by the discrepancy between the physical and analytic currents on the boundary, leading to modest changes in plasma heating [Fig.~\ref{fig:j2_energy}(b)].

When $\mathbf{E_1}$, $\mathbf{B_1}$, and $\mathbf{J_1}$ are not a good solution to Maxwell's equations, the system of equations solved in the simulation is modified. For example, if $\mathbf{J_1} = \mathbf{J_\mathbf{1,0}} + \Delta \mathbf{J}$, where $\mathbf{J_\mathbf{1,0}}$ solves Maxwell's equations exactly along with $\mathbf{E_1}$ and $\mathbf{B_1}$, then the simulation will function as if it contains an extra current source term corresponding to $\Delta \mathbf{J}$. This source term may be fairly benign (if $\Delta \mathbf{J}$ is non-resonant and small), but it can also be significantly disruptive (when $\Delta \mathbf{J}$ is not very small and drives the plasma near resonance). 
As a result, a loss of energy conservation may be observed when Eqs.~\ref{eqn:e_amp_j},~\ref{eqn:j1}, and~\ref{eqn:b_corr_t} do not provide an accurate solution to Maxwell's equations and the plasma is driven close to its resonant frequency for a substantial amount of time.
This occurs, for example, when the finite-duration correction is not included, even if the pulse duration is fairly long [no FD case in Fig.~\ref{fig:j2_energy}(a)]. Loss of energy conservation also occurs for long-distance propagation of non-relativistic, sufficiently short pulses in sufficiently dense plasma [e.g., the nominal $n_\mathrm{e} = 0.05 \, n_\mathrm{c}$ cases in Fig.~\ref{fig:j2_energy}(c) and~\ref{fig:j2_energy}(d)], where higher-order finite-duration corrections may be required for stability.

For short pulses, energy conservation may be recovered by lowering the plasma density [$0.01\,n_\mathrm{c}$ case in Fig.~\ref{fig:j2_energy}(c)] or detuning the analytic terms from resonance with the plasma response. The latter is naturally produced by relativistic modification of the phase velocity [as in the $a_0=1$ case in Fig.~\ref{fig:j2_energy}(c)], which makes APT with analytic current suitable for modeling relativistic pulses regardless of duration. However, because the validity of the vacuum boundary condition requires low intensity at the domain boundary, relativistic pulses can only be accurately modeled when focusing is included. The unique capabilities of APT, such as capturing focusing regardless of simulation dimensionality and the ability to model relativistic pulses in plasma-filled domains, may thereby be combined to enable novel simulation configurations.

\section{Conclusions}

In summary, a new approach to computational electromangetics that splits the underlying physics equations into analytic and computational parts overcomes limitations of conventional grid-based methods. 
This analytic pulse technique (APT) introduces additional flexibility while maintaining physical fidelity. 
APT can incorporate the electric and magnetic fields of an approximate solution to Maxwell's equations into an otherwise self-consistent simulation, allowing the validity of the approximation to be evaluated directly from the observed material response.
This validation procedure bypasses the conventional constraint on dimensionality, e.g., allowing a 3-D--like solution to be tested in a series of non-interacting 1-D simulations.
The use of analytic fields and material response terms can also simplify the numerical implementation of boundary conditions and reduce or eliminate related errors.
The combination of these properties enables a new simulation paradigm: using APT with validated approximate solutions to Maxwell's equations can allow simulations to be constrained in size and dimensionality by the material response rather than the need to correctly inject and evolve electromagnetic waves. 
As a result, APT can new open areas to investigation that have traditionally been inaccessible for reasons of numerics, computational cost, or algorithmic complexity.

In the context of particle-in-cell simulation of laser--plasma interaction, APT was verified to produce the the same dynamics as the conventional approach when the analytic pulse and optional, analytic current represent solutions to Maxwell's equations including numerical dispersion. Direct evaluation of theoretically relevant, approximate solutions to Maxwell's equations, such as Laguerre--Gaussian beams, is enabled by relaxation of the ordinary requirement for the laser pulse to be a full solution to Maxwell's equations. APT also allows the effects of 3-D focusing on a laser pulse to be included in lower-dimensional simulations and simplifies the inclusion of novel laser pulses with desirable but otherwise prohibitively expensive or complex focal characteristics. When analytic currents are added, APT can simplify boundary conditions for plasma-filled domains, facilitating the long-distance modeling of focusing laser pulses in low-density plasma. In addition, APT is cost-effective in particle-in-cell simulation. The analytic pulse technique is thereby expected to facilitate the study of next-generation laser--plasma problems.

Future work will focus on leveraging the unique capabilities of the analytic pulse technique. 
For example, tightly focused, ultrahigh-intensity laser pulses are of significant interest for QED problems including electron--positron pair cascades~\cite{sokolov2010compton,bulanov2013compton,jirka2017pair}. However, the inaccuracy of analytic descriptions away from focus, where pulses must be introduced using traditional methods, requires custom pulse injection techniques~\cite{li2016tight,thiele2016tight,jirka2017pair}.
APT shifts the burden of accuracy to the highest-intensity region, where pulses are well-described, and will facilitate modeling in this regime.
The reductions in simulation dimensionality and cost afforded by APT are also ideal for machine-learning--based optimization of QED observables.
In addition, the use of analytic pulses is of significant interest to model previously inaccessible configurations in the long-distance interaction of flying-focus pulses with low-density plasma, where control of the intensity dynamics is desirable to drive physical processes such as wakefield formation~\cite{palastro2020dephasingless} and radiation generation~\cite{ramsey2022nlts}.

\section*{Acknowledgments}

This material is based upon work supported by the Department of Energy National Nuclear Security Administration under Award Number DE-NA0003856, the Department of Energy Office of Fusion Energy Sciences under Award Number DE-SC0021057, the University of Rochester, and the New York State Energy Research and Development Authority.
The support of DOE does not constitute an endorsement by DOE of the views expressed in this paper.
The work of M.V. is supported by the Portuguese Science Foundation (FCT) Grant No. CEECIND/01906/2018 and PTDC/FIS-PLA/3800/2021.
Particle-in-cell simulations were performed using \textit{EPOCH},\cite{arber2015epoch} developed under UK EPSRC Grant Nos. EP/G054940, EP/G055165, and EP/G056803. 
This work used Stampede2 at the Texas Advanced Computing Center through allocation TG-PHY210072 from the Advanced Cyberinfrastructure Coordination Ecosystem: Services \& Support (ACCESS) program, which is supported by National Science Foundation grants \#2138259, \#2138286, \#2138307, \#2137603, and \#2138296.
 
This report was prepared as an account of work sponsored by an agency of the U.S. Government. Neither the U.S. Government nor any agency thereof, nor any of their employees, makes any warranty, express or implied, or assumes any legal liability or responsibility for the accuracy, completeness, or usefulness of any information, apparatus, product, or process disclosed, or represents that its use would not infringe privately owned rights. Reference herein to any specific commercial product, process, or service by trade name, trademark, manufacturer, or otherwise does not necessarily constitute or imply its endorsement, recommendation, or favoring by the U.S. Government or any agency thereof. The views and opinions of authors expressed herein do not necessarily state or reflect those of the U.S. Government or any agency thereof.

\appendix

\section{Validity of the small frequency bandwidth approximation} \label{append:small}

The approximate analytic expression for a finite-duration plane-wave pulse composed of a non-evolving envelope multiplied by a phase remains valid over only a finite time in the presence of phase velocity dispersion. Consider the phase of two frequency components in a pulse separated by $\Delta \omega$. At time $t$ and position $x$, the phases of the frequency components are given by
\begin{align}
& \phi_\omega = \omega \left[t - \dfrac{x}{v_\phi\left(\omega\right)}\right] \\
& \begin{aligned} \phi_{\omega+\Delta\omega} & = \left(\omega+\Delta\omega\right)\left[t - \dfrac{x}{v_\phi \left(\omega+\Delta\omega\right)}\right] \\
& \approx \left(\omega+\Delta\omega\right)\left[t - \left(1 - \dfrac{\Delta\omega}{v_\phi\left(\omega\right)}\left.\dd{v_\phi}{\omega}\right|_\omega \right)\dfrac{x}{v_\phi \left(\omega\right)}\right],
\end{aligned}
\end{align}
where the last expression was obtained through Taylor expansion of $v_\phi(\omega+\Delta\omega)$, assuming $\Delta \omega \ll \omega$. At a later time $t + t_1$, the frequency component $\omega$ has the same phase at position $x + v_\phi(\omega)t_1$, where the phase of the second frequency component has changed by
\begin{equation} \label{eqn:mono}
    \Delta \phi_{\omega+\Delta\omega} \approx \dfrac{\omega}{v_\phi\left(\omega\right)}\left.\dd{v_\phi}{\omega}\right|_\omega \Delta\omega\, t_1.
\end{equation}

In vacuum, $\mathrm{d}v_\phi/\mathrm{d}\omega$ is purely numerical in origin can be neglected when the simulation is adequately resolved.
Assuming the correction to the phase velocity is $n$th-order in $\omega \Delta x/c$, (e.g., $n=2$, as in Eq.~\ref{eqn:vphi}), the phase shift is to lowest order
\begin{equation}
    \Delta \phi_{\omega+\Delta\omega} \approx - \left[1-\dfrac{v_\phi\left(\omega\right)}{c}\right] n \Delta \omega\, t_1,
\end{equation}
where the term in brackets is nonzero only due to resolution-related effects.
In terms of the phase shift in the $\omega$ component relative to physical, dispersionless vacuum ($\Delta \phi_{\omega,\,\mathrm{vac}}$),
\begin{equation}
\Delta \phi_{\omega+\Delta\omega} \approx - \dfrac{n \Delta \omega}{\omega} \Delta \phi_{\omega,\,\mathrm{vac}}.
\end{equation}
Provided the pulse has a sufficient number of cycles to satisfy $n \Delta \omega/\omega \ll 1$, conditions which deliver reasonable accuracy in the phase of the central frequency of the pulse will also thereby be guaranteed to maintain the validity of Eq.~\ref{eqn:plane_func} in vacuum.

In plasma, however, dispersion may also be physical in origin. In the case where analytic currents are included, the bandwidth of the pulse must be sufficiently narrow not to incur significant phase slip due to the physical dependence of $v_\phi$ on $\omega$. When physical dispersion dominates over numerical dispersion, the requirement for the small-bandwidth approximation to be valid ($\Delta \phi_{\omega+\Delta\omega} \ll 1$ in Eq.~\ref{eqn:mono}) is 
\begin{equation}
\dfrac{\omega_p^2/\omega^2}{1-\omega_p^2/\omega^2} \Delta\omega\, t \ll 1.
\end{equation}
Assuming a temporal feature of duration $\tau$ in the pulse sets the bandwidth as $\Delta \omega \sim 1/\tau$, the time for which the small-bandwidth approximation is valid is therefore
\begin{equation}
t \lesssim \left(\dfrac{\omega^2}{\omega_p^2}-1\right) \tau.
\end{equation}
Longer simulations may require accounting for bandwidth, i.e., specifying the electric field in Eq.~\ref{eqn:mono} via Fourier transform with frequency-dependent expressions for $v_{\mathrm{g}}$ and $v_\phi$, in order to incorporate the effect of group velocity dispersion. This calculation and its relevance for the second-order explicit Yee scheme are discussed in Appendix~\ref{append:dispersion}.

\section{Effect of the incoming wave boundary condition on the pulse amplitude} \label{append:boundary}

A normally incident plane wave is launched from the simulation boundary using Ampere's law and the incoming wave boundary condition. For the minimum-$x$ boundary and $y$ polarization, these equations are
\begin{align}
& \pd{B_z}{x} = - \dfrac{1}{c}\pd{E_y}{t} + \pd{B_x}{z} - J_y \\
& E_y + B_z = 2S + E_y^* + B_z^*,
\end{align}
where $E^*$ and $B^*$ are the initial values at $t=0$, and $S$ is the source term. In \textit{EPOCH}, $E$ is set on the edge of the simulation domain (subscript $\mathrm{b}$) by setting the value of $B$ a half-step outside the domain (subscript $-1/2$).
Omitting $B_x$ and $J_y$, these equations are discretized as
\begin{align}
    & B_{1/2}^{n+1/2}-B_{-1/2}^{n+1/2} = - \dfrac{\Delta x}{c\Delta t} \left(E_\mathrm{b}^{n+1}-E_\mathrm{b}^{n}\right) \\
    & E_\mathrm{b}^{n+1} + E_\mathrm{b}^{n} + B_{-1/2}^{n+1/2} + B_{1/2}^{n+1/2} = 4 S_\mathrm{b}^{n+1/2}.
\end{align}
These equations determine the amplitude of the electromagnetic wave that propagates into the domain, setting $E_0$ through $E_\mathrm{b}$.
The solution for a plane wave gives
\begin{equation}
    \begin{aligned}
    S_0  & = \dfrac{1}{2} E_0 \left[ \cos\left(\omega \Delta t/2\right) + \dfrac{\Delta x}{c\Delta t} \dfrac{\sin \left(\omega \Delta t/2\right) }{\sin \left(k \Delta x/2\right)} \cos \left(k \Delta x/2\right) \right] \\
    & \approx \left(1 - \dfrac{1}{16}\left[1+\left(\dfrac{c\Delta t}{\Delta x}\right)^2\right]\left[\dfrac{\omega\Delta x}{c}\right]^2\right) E_0.
    \end{aligned}
\end{equation}

In addition to the amplitude, the boundary condition also modifies the envelope of the pulse due to the frequency dependence of $S_0$. Also, while it is convenient to specify the pulse envelope in terms of $E$ in the analytic pulse technique, the source term in the boundary condition used in \textit{EPOCH} is more closely linked with $B$. Therefore, even though laser pulses in the analytic pulse technique and all-internal simulations both represent valid solutions to Maxwell's equations, the pulse envelope is expected to differ by an amount comparable to the difference between $E$ and $B$ (i.e., the finite-duration correction). In vacuum, the difference in the envelopes of $E$ and $B$ is purely numerical and vanishes for infinitely well-resolved simulations, leading to the second-order scaling in Fig.~\ref{fig:demo}(e). In plasma, however, the difference in the envelopes of $E$ and $B$ is physical and does not vanish in the infinitely resolved case.

\section{Approximate dispersive corrections} \label{append:dispersion}

Consider a laser pulse defined through Fourier transform by
\begin{equation}
E = \int \Tilde{E}\left(\omega\right)\exp{\left(i\omega t-ikx\right)}\mathrm{d}\omega,  
\end{equation}
where $k = \omega/v_\phi(\omega)$ satisfies the dispersion relation. Assuming $\Tilde{E}$ is peaked around a central frequency $\omega_0$, $k$ can be written as a Taylor expansion about $\omega_0$,
\begin{equation}
k \approx \dfrac{\omega_0}{v_{\phi,0}} + \dfrac{\left(\omega - \omega_0\right)}{v_{\mathrm{g},0}} - \dfrac{\left(\omega - \omega_0\right)^2v_{\mathrm{g}}'}{2 v_{\mathrm{g},0}^2}
\end{equation}
where the GVD term $v'_{\mathrm{g}} = \mathrm{d}v_{\mathrm{g}}/\mathrm{d}\omega$ is evaluated at $\omega_0$.
The electric field is then given by
\begin{equation}
E \approx \exp{\left[i\omega_0\left(t-x/v_{\phi,0}\right)\right]} \int \Tilde{E}\left(\omega\right)\exp{\left[i\left(\omega-\omega_0\right)\left(t-x/v_{\mathrm{g},0}\right) +i\alpha \left(\omega-\omega_0\right)^2 \right]}\mathrm{d}\omega,
\end{equation}
where
\begin{equation}
    \alpha \equiv \dfrac{v_{\mathrm{g}}'x}{2 v_{\mathrm{g},0}^2}.
\end{equation}
In the limit that the correction factor is small, $\exp[i\alpha (\omega-\omega_0)^2]\approx 1 + i\alpha(\omega-\omega_0)^2 - \alpha^2(\omega-\omega_0)^4/2$ and the integral becomes
\begin{equation} \label{eqn:chromatic_corr}
\begin{aligned}
\int \Tilde{E} & \left(\omega\right)\exp{\left[i\left(\omega-\omega_0\right)\left(t-x/v_{\mathrm{g},0}\right) +i\alpha \left(\omega-\omega_0\right)^2 \right]}\mathrm{d}\omega \\
& \approx \int \Tilde{E}\left(\omega\right)\exp{\left[i\left(\omega-\omega_0\right)\left(t-x/v_{\mathrm{g},0}\right)\right]}\left[1 + i\alpha \left(\omega-\omega_0\right)^2 - \dfrac{1}{2}\alpha^2\left(\omega-\omega_0\right)^4\right]\mathrm{d}\omega.
\end{aligned}
\end{equation}
The first term is the usual monochromatic result, while the second and third terms represent the lowest-order effects of GVD on phase and amplitude, respectively. 
Comparison with the originally monochromatic pulse defined by Eq.~\ref{eqn:e_amp} indicates that $\Gamma$ and $\Phi$ of the envelope are modified to new values $\Gamma_\alpha$ and $\Phi_\alpha$.
Consider a pulse envelope given by $\Gamma_0 e^{i\Phi_0}$ in the limit that $\alpha \to 0$ (i.e., at small $x$). Let $\Gamma_0$ and $\Phi_0$ be functions of the pulse-envelope coordinate $\eta = (t-x/v_{\mathrm{g},0})/\tau $, where $\tau$ is a constant.
The lowest-order GVD correction can be approximated from Eq.~\ref{eqn:chromatic_corr} using the derivative property of the Fourier transform,
\begin{equation}
    \Gamma_\alpha \left(\eta\right) e^{i\Phi_\alpha \left(\eta\right)} \approx \Gamma_0 e^{i\Phi_0} -i\dfrac{\alpha}{\tau^2} \dd{^2\left(\Gamma_0 e^{i\Phi_0}\right)}{\eta^2} - \dfrac{\alpha^2}{2\tau^4} \dd{^4 \left(\Gamma_0 e^{i\Phi_0}\right)}{\eta^4}.
\end{equation}
To second order in $\alpha$, this gives 
\begin{align}
& \begin{aligned}
    \Gamma_\alpha \approx \Gamma_0 & \left( 1 + \dfrac{\alpha}{\tau^2}\left[ \dfrac{2 \Gamma_0' \Phi_0'}{\Gamma_0} +\Phi_0'' \right] \right. \\
     & \phantom{+} \left. + \dfrac{\alpha^2}{2\tau^4}\left[ - \dfrac{\Gamma_0^{(4)}}{\Gamma_0} + \left(\dfrac{\Gamma_0''}{\Gamma_0}\right)^2 +\dfrac{4 \Gamma_0'' \left(\Phi_0'\right)^2}{\Gamma_0} + \dfrac{12 \Gamma_0' \Phi_0' \Phi_0''}{\Gamma_0} + 3 \left(\Phi_0''\right)^2 + 4 \Phi_0' \Phi_0''' \right] \right)
\end{aligned} \\
    & \Phi_\alpha \approx \Phi_0 + \dfrac{\alpha}{\tau^2} \left[ -\dfrac{\Gamma_0''}{\Gamma_0} + \left(\Phi_0'\right)^2 \right] - \dfrac{2\alpha^2}{\tau^4} \left[ \dfrac{\Gamma_0'''\Phi_0'}{\Gamma_0}+\dfrac{\Gamma_0''\Phi_0''}{\Gamma_0}+\dfrac{\Gamma_0'\Phi_0'''}{\Gamma_0}-\left(\Phi_0'\right)^2\Phi_0''-\dfrac{1}{4}\Phi_0^{(4)}\right],
\end{align}
where primes denote derivatives in $\eta$.
The current and the magnetic field can then be found from the electric field using Eqs.~\ref{eqn:j1} and~\ref{eqn:b_corr_t}, respectively, with $\Gamma \to \Gamma_\alpha$ and $\Phi \to \Phi_\alpha$.

This approach can be extended to arbitrary order by retaining more terms in the Taylor expansions, but is restricted to $\alpha/\tau^2 < 1$.
In the case with analytic currents, and assuming physical dispersion dominates over numerical dispersion, this GVD correction enables simulation in low-density plasma ($\omega_p \ll \omega_0$) for distances
\begin{equation}
    x \lesssim \dfrac{2 \omega_0^3 c \tau^2}{\omega_p^2},
\end{equation}
which is a factor of $\sim \omega_0 \tau$ increase over the uncorrected limit.

The application of GVD corrections, however, is only physically justifiable when the GVD contribution to the phase and envelope dynamics dominates over the numerical effect of imperfect dispersion. Consider, for example, the GVD phase correction and the numerical contribution to the phase from imperfect dispersion, both of which increase linearly with propagation distance. For an $n$th-order scheme, the ratio of these contributions is approximately
\begin{equation}
\dfrac{\Delta \phi_n}{\Delta \phi_\mathrm{GVD}} \approx 2 f_n \dfrac{\omega \tau^2  v_{\mathrm{g}}^2}{v_{\mathrm{g}}' v_\phi} \left(\dfrac{\omega \Delta x}{c}\right)^n  \approx 2 f_n \left(1-\omega_p^2/\omega^2\right)^2 \omega^2 \tau^2  \dfrac{\omega^2 }{\omega_p^2} \left(\dfrac{\omega \Delta x}{c}\right)^n   ,
\end{equation}
where $f_n (\omega \Delta x/c)^n$ is the difference between the physical and numerical phase velocities, $(1/\Gamma)(\mathrm{d}^2\Gamma/\mathrm{d}\eta^2)\sim 1$ was used, and the second expression assumes the phase and group velocities are dominated by the plasma density rather than numerical effects. With the second-order explicit Yee scheme, it is usually the case in underdense plasma that $\Delta \phi_n \gtrsim \Delta \phi_{\mathrm{GVD}}$ (e.g., for the parameters used in the present work) unless very high resolution is employed. In other words, the second-order explicit Yee scheme is not well-suited to modeling the regime in which group velocity dispersion is important. Correction terms are more likely to be applicable for higher-order solvers, which are a better choice for systems that are sensitive to GVD.

\section*{Data Availability}

The data that support the findings of this study are available from the corresponding author upon reasonable request. 

\section*{References}


\end{document}